\documentclass{emulateapj}

\defcitealias{My2006ApJ}{Paper~I}

\slugcomment{Accepted for publication in ApJ}

\usepackage{ulem}
\usepackage{color}

\shorttitle{STAR FORMATION IN THE EXTREME OUTER GALAXY}

\shortauthors{Chikako Yasui et al.}

\begin{document}

\title{STAR FORMATION IN THE EXTREME OUTER GALAXY: \\ DIGEL CLOUD 2
CLUSTERS\footnote{Based on data collected at Subaru Telescope, which is
operated by the National Astronomical Observatory of Japan.}}

\author{Chikako Yasui and Naoto Kobayashi\footnote{Also at: Subaru
Telescope, National Astronomical Observatory of Japan, 650 North A`ohoku
Place, Hilo, HI 96720, USA.}}
\affil{Institute of Astronomy, School of Science, University of Tokyo,
2-21-1 Osawa, Mitaka, Tokyo 181-0015, Japan}
\email{ck\_yasui@ioa.s.u-tokyo.ac.jp}

\author{Alan T. Tokunaga}
\affil{Institute for Astronomy, University of Hawaii, 2680 Woodlawn
Drive, Honolulu, HI 96822, USA}

\author{Hiroshi Terada} 
\affil{Subaru Telescope, National Astronomical Observatory of Japan, 650
North A`ohoku Place, Hilo, HI 96720, USA}

\and

\author{Masao Saito} 
\affil{ALMA Project, National Astronomical Observatory of Japan, 2-21-1
Osawa, Mitaka, Tokyo 181-8588, Japan}


\begin{abstract}

As a first step for studying star formation in the extreme outer Galaxy
(EOG), we obtained deep near-infrared ($J, H, K$-bands) images of two
embedded clusters at the northern and southern CO peaks of Cloud 2,
which is one of the most distant star forming regions in the outer
Galaxy (galactic radius $R_g \sim 19$ kpc).  With high spatial
resolution (FWHM $\sim 0\farcs3$--$0\farcs35$) and deep imaging ($K \sim
21$ mag, 5 $\sigma$) with the IRCS imager at the Subaru telescope, we
detected cluster members with a mass detection limit of $< 0.1 M_\odot$,
which is well into the substellar regime.  These high quality data
enables a comparison of EOG to those in the solar neighborhood on the
same basis for the first time.  Before interpreting the photometric
result, we have first constructed the NIR color-color diagram (dwarf
star track, classical T Tauri star (CTTS) locus, reddening law) in the
Mauna Kea Observatory filter system and also for the low metallicity
environment since the metallicity in EOG is much lower than those in the
solar neighborhood. The estimated stellar density suggests that an
``isolated type'' star formation is ongoing in Cloud 2-N, while a
``cluster type'' star formation is ongoing in Cloud 2-S.
Despite the difference of the star formation mode, other characteristics
of the two clusters are found to be almost identical: (1) $K$-band
luminosity function (KLF) of the two clusters are quite similar, as is
the estimated IMF and ages ($\sim 0.5$--1 Myr) from the KLF fitting,
(2) the estimated star formation efficiencies (SFEs) for both clusters
    are typical compared to those of embedded clusters in the solar
    neighborhood ($\sim 10$ \%).
The similarity of two independent clusters with a large separation
 ($\sim 25$ pc) strongly suggest that their star formation activities
 were triggered by the same mechanism, probably the supernova remnant
 (GSH 138-01-94) as suggested in Kobayashi et al. and Yasui et al.

\end{abstract}



\section{Introduction}
\label{sec:intro}

The extreme outer Galaxy (hereafter ``EOG''), defined here as the region
at Galactic radius ($R_g$) greater than 18 kpc, could be regarded as the
outermost region of the Galaxy because the distribution limits of
Population I and II stars are known to be 18--19 kpc and 14 kpc,
respectively \citep{Digel1994}. The EOG is known to have very low gas
density \citep{{BW95},{Nakanishi03}} and low metallicity
\citep{Smartt1997}. Such an extreme environment could serve as a
``laboratory'' for studying the star formation process in a totally
different environment from that in the solar neighborhood.
The star-forming process in such environments has not been studied
extensively as for the nearby star-forming regions because of the large
distance. The EOG does not seem to have long and complex star formation
history because there is little or no perturbation from the spiral arms,
which is the major and powerful trigger for continuously converting a
large amount of molecular gas into stars.
Therefore, in the EOG there is a good chance to observe the detail of
the each star formation process, such as the triggering process of
cluster formation, without being disturbed by the entangled star
formation history in space and time.

%
There are some synthetic studies of molecular clouds in EOG and far
outer Galaxy (hereafter ``FOG''), whose $R_g$ is not more than 18 kpc
but no less than 15 kpc. The pioneering survey of molecular clouds and
star formation beyond the solar circle was undertaken by \citet{WB89},
who identified the $^{12}$CO molecular emission associated with
color-selected IRAS point sources and used kinematic distances to place
the sources within the Galaxy. Many of the IRAS sources were found to
lie in the far outer Galaxy. Follow-up studies of the gas properties of
many of these far outer Galaxy molecular clouds were presented in
\citet{{BW94},{BW95}} separately.  In an independent search,
\citet{Digel1994} found eleven molecular clouds in the EOG, based on CO
observations of distant \ion{H}{1} peaks in the Maryland-Green Bank
survey \citep{WW1982}. Following those pioneering works, \citet{Heyer98}
conducted a comprehensive CO survey of molecular clouds in the FOG with
Five College Radio Astronomy Observatory (FCRAO) CO survey.  With
H$\alpha$ and radio continuum study, \citet{Fich84}, \citet{BrandB93},
and \citet{Rudolph96} identified a number of \ion{H}{2} regions
associated with the FOG clouds, indicating the presence of recent
massive star formation.

%
For the study of star formation, it is ultimately necessary to observe
YSOs (young stellar objects) at near infrared (NIR) wavelengths.
\citet{KT2000} found YSOs associated with a molecular cloud in EOG for
the first time with a wide-field NIR survey. This cloud is denoted as
Cloud 2 in the list of \citet{Digel1994} and is one of the most distant
molecular clouds in the outer Galaxy region.  It is located at a very
large Galactic radius ($R_g = 18$--28 kpc) in the second quadrant of the
Galaxy ($l \sim 137.75^\circ$, $b \sim -1.0$).
It is by far the largest molecular cloud among those found in EOG with a
mass comparable to that giant molecular clouds ($M \sim 2 \times 10^4$
$M_\odot$; Digel et al. 1994).
The Galactic radius of Cloud 2 has been estimated by various methods at
$R_g = 28$ kpc (heliocentric distance: $D=21$ kpc) from the kinematic
distance of the CO cloud \citep{Digel1994}; $R_g = 23.6$ kpc ($D=16.6$
kpc) from the latest \ion{H}{1} observation \citep{Stil2001}; and $R_g =
15$--19 kpc ($D=8.2$--12 kpc) from the detailed optical study of a
B-type star, MR1 \citep{{Muzzio1974},{Smartt1996}}, which is associated
with Cloud 2 \citep{deGeus93}.  Throughout this paper we adopt $R_g = 19$
kpc ($D=12$ kpc) because it is about the mean value of all the estimated
distances and because spectroscopic distance of stars should be more
accurate than kinematic distance.
Metallicity at the Galactic radius of Cloud 2 is estimated to be $\sim
-1$ dex from the standard metallicity curve by \citet{Smartt1997}. In
fact, the metallicity of the B-type star, MR1, is measured at $-0.5$ to
$-0.8$ dex by optical high-resolution spectroscopy
\citep{{Smartt1996},{Rolleston2000}} and that of the Cloud 2 itself is
measured at $-0.7$ dex from the radio molecular emission lines
\citep{{Lubowich2004}, {Ruffle07}}. These metallicity values are
comparable to that of LMC ($\sim -0.5$ dex) and SMC \citep[$\sim -0.9$
dex;][]{Arnault}.
Using deeper near-infrared imaging with QUIRC at the University of
Hawaii (UH) 2.2m telescope, Kobayashi et al. (2007) also found two
embedded clusters at the northern and southern CO peaks of Cloud 2.
They suggested that the overall star formation activity in Cloud 2 was
triggered by the supernova remnant (SNR) \ion{H}{1} shell, GSH
138-01-94, and that supernova triggered star formation could be one of
the major star formation modes in EOG.

Recently Brand \& Wouterloot (2007) reported comprehensive study of a
star forming region, WB89-789, whose distance is estimated at $D = 11.9$
kpc, $R_g \sim 20.2$ kpc. This region was found in \citet{WB89}
described above.  They obtained NIR imaging with limiting magnitude of
$m_K = 16.5$ and 17.5 mag for 10 $\sigma$ and 5 $\sigma$, respectively,
which correspond to a 2--3 $M_\odot$ dwarf main sequence star with 10
mag of visual extinction.  They also used molecular line and dust
continuum observations for properties of the molecular cloud and showing
the existence of an outflow.  They also obtained optical spectroscopy of
a star for confirmation of the distance to the region.
\citet{Santos2000} discovered two distant embedded young clusters in the
FOG ($R_g = 16.5$ kpc; $D=10.2$ kpc) with near-infrared imaging of CO
clouds associated with IRAS sources. Their detection limit ($m_K = 16.4$
mag) corresponds to a 1 Myr old star of $\sim$1--2 $M_\odot$ seen
through 10 mag of visual extinction.
\citet{Snell2002} have conducted a comprehensive NIR survey of embedded
clusters in the FOG ($R_g = 13.5$--17.3 kpc) based on the FCRAO CO
survey \citep{Heyer98}. They found 11 embedded clusters with the
detection limit of $m_{K'}$=17.5 mag. For their most distant cluster,
this magnitude corresponds to a 1 Myr old, 0.6 $M_\odot$ star with no
extinction, or a 1 Myr old, 1.3 $M_\odot$ star seen through 10 mag of
visual extinction.
These independent near-infrared searches so far found cluster members
with the mass detection limit of $\sim 1$ $M_\odot$.

As a next step to study the details of the star formation activity in
the EOG, we obtained deep and high spatial resolution NIR images of the
two most distant embedded clusters in Digel Cloud 2 with Subaru 8.2-m
telescope.  Because of the achieved sensitivity to detect substellar
mass members,
we could compare embedded clusters in EOG to those in the solar
neighborhood for the first time.
In \citet[][hereafter Paper~I]{My2006ApJ}, we presented the data for the
Cloud 2-N cluster and (i) estimated the age of the cluster at less than
1 Myr by comparing the observed KLF to model KLFs based on typical IMFs,
and (ii) suggested that the formation of the cluster was triggered by
the SNR shell based on the age and the geometry of the cluster.  In this
paper, we present data of the Cloud 2-S cluster and also re-analyze the
data of Cloud 2-N cluster to investigate the fainter stellar population.
For interpreting the photometric result, we have constructed
for the first time the NIR color-color diagram (dwarf track, CTTS locus,
and reddening vector) in the Mauna Kea Observatory (MKO) near-infrared
filter system \citep{{Simons2002},{Tokunaga2002}}.
To accommodate to the low-metallicity in EOG, we also considered the
mass-luminosity relation and the color-color diagram for low metallicity
stars.  The established methods here would be useful for future study of
star formation in EOG and FOG, and ultimately nearby dwarf galaxies.
We then discuss the KLF and star formation efficiency in the EOG based
on the results of these two clusters.


\section{Observations and Data Reduction} \label{OBSandDATA}

\subsection{Subaru IRCS imaging} \label{IMAGE}

We obtained $J$ (1.25 $\mu$m), $H$ (1.65 $\mu$m), $K$ (2.2 $\mu$m)-band
deep images of the Cloud 2-N and -S clusters
with the Subaru Infrared Camera and Spectrograph
(IRCS; Tokunaga et al. 1998; Kobayashi et al. 2000; Terada et al. 2004)
with a pixel scale of $0\farcs058$ pixel$^{-1}$.
IRCS employs the Mauna Kea Observatory (MKO) near-infrared photometric
filters \citep{{Simons2002},{Tokunaga2002}}.
The entire infrared cluster was sufficiently covered with the $\sim 1'$
field of view.
For the Cloud 2-N cluster, the observation was conducted on 2000 December
1 UT with the total integration time of 600, 450, 900 s for $J$, $H$,
and $K$ bands, respectively.
For the Cloud 2-S cluster, the observation was conducted on 2000 October
25 UT with the total integration time of 600, 600, 420 s for $J$, $H$,
and $K$ bands, respectively.
The observing condition was photometric and the seeing was excellent
($\sim 0\farcs3$--$0\farcs35$) throughout the observing period on both
nights.

\subsection{Data Reduction} \label{sec:Data}

All the data for each band were reduced with IRAF\footnote{IRAF is
distributed by the National Optical Astronomy Observatories, which are
operated by the Association of Universities for Research in Astronomy,
Inc., under cooperative agreement with the National Science Foundation.}
with standard procedures: dark subtraction, flat fielding, bad-pixel
correction, median-sky subtraction, image shifts with dithering offsets,
and combining. 
In our previous paper \citepalias{My2006ApJ}, we used the DAOFIND
algorithm in IRAF DAOPHOT package with the detection threshold of 10
$\sigma$ for the source detection in the Cloud 2-N cluster.  In the
present study, we performed
source detection for the Cloud 2-N cluster again and the Cloud 2-S
cluster using SExtractor (Bertin et al. 1996) with the detection
threshold of 5 $\sigma$ above the local background. 
Photometry was performed using IRAF APPHOT package with aperture
diameters of $0\farcs58$ (10 pixel) and $0\farcs46$ (8 pixel), for Cloud
2-N and -S, respectively. The aperture sizes were chosen to achieve the
highest S/N.
For four pairs of binary stars in the Cloud
2-N cluster, $0\farcs 35$ (6 pix) diameter aperture was applied with an
aperture correction because the separation is less than $0\farcs 5$.
For photometric calibration we used four and seven bright and isolated
stars for the Cloud 2-N and -S cluster, respectively, in the images for
which photometry was accurately done later with newly obtained $JHK$
images of the fields with Subaru's new wide-field near-infrared camera
MOIRCS (Ichikawa et al. 2006).  The flux uncertainties are estimated as
the same way as in Paper I.  The resultant limiting magnitudes (5
$\sigma$) for the Cloud 2-N cluster are $J = 22.0$, $H = 21.0$, and $K =
20.9$, while those for the Cloud 2-S cluster are $J=21.8$, $H=21.0$, and
$K=20.4$.  The pseudocolor images of the observed field and the spatial
distributions of detected stars are shown in Figs.~1 and 2 for the Cloud
2-N cluster and -S cluster, respectively.


\section{Color-Color Diagram in MKO Photometric System}
\label{sec:colMKO}

For identifying of young cluster members, $JHK$ color-color diagram with
dwarf and giant star tracks in Johnson-Glass system \citep[][hereafter
``B\&B'']{Bessell}, CTTS locus in CIT system \citep{T Tauri}, and
reddening vector in Arizona system \citep{RL} have been widely used
\citep[e.g.,][]{Lada00AJ}.
Because the Mauna Kea Observatory (MKO) filter system
\citep{{Simons2002},{Tokunaga2002}}, which we used for the present
study, is relatively new, the color-color diagram in this system have
not yet been clearly established.
Although we used the color-color diagram with the dwarf track etc. in
other systems in our previous paper \citepalias{My2006ApJ},
we constructed a color-color diagram solely in the MKO system for more
rigorous treatment of the data.

Firstly, we plotted colors of dwarf stars\footnote{Plotted are only for
dwarfs earlier than M6, which is the end point of the B\&B dwarf track.}
observed with MKO filters
%
(Hewett et al. 2006, open circle, B9V--M6V\footnote{\citet{Hewett2006}
shows B--Y dwarf stars. These are computed from their own
spectra. (synthetic colors)};
%
%
%
Leggett et al. 2002, plus, M4V--M6V\footnote{\citet{Leggett2002} shows
the colors of M4V--T8V stars.};
%
%
%
%
Leggett et al. 2006, cross, M4--M6\footnote{\citet{Leggett2006}
shows the colors of B7.5--M9.5 stars. We plotted dwarf stars and also
stars without clear designation of luminosity class.})
%
%
to construct a dwarf track in the MKO system (solid line).
Although this MKO track appears to be almost identical to the B\&B track
for stars earlier than K type stars, the MKO track has a smaller $J-H$
color than the B\&B track for M type stars.
Because the MKO $J$ filter has a narrower band width compared to the
other filter systems \citep[see Fig.~5 in][]{filters}, the MKO $J$
magnitude is less affected by the strong H$_2$O vapor absorption bands
of M type stars. This should result in a brighter $J$ magnitude, thus
the smaller $J-H$ color in the MKO system. We interpret this as the
reason for the shift of the track position for M type stars and also the
reason why the shift becomes larger with later spectral type.

%
Secondly, in order to construct the CTTS locus in the MKO system, we
checked the 2MASS $J, H, K$-band magnitudes of
classical T Tauri stars\footnote{Although \citet{T Tauri} used 30 CTTSs
from Strom et al. (1989) for their derivation, we found 33 usable CTTSs
with the spectral type range of G0--M5 as defined in Strom et
al. (1989). We removed four stars from our analysis because their 2MASS
photometric quality is not good. }
in Strom et al. (1989), which were used to derive the original CTTS
locus in the CIT system \citet{T Tauri}.
After dereddening the 2MASS colors of these stars following the method
of \citet{T Tauri} with the reddening law in \citet{2MASSav}, we
converted the 2MASS colors to the MKO colors using the conversion in
\citet{Leggett2006}.
We derived the CTTS locus in MKO system by least-square fitting of the
MKO colors: the resultant locus is $(J-H)_{\rm MKO} = (0.32 \pm 0.07)
(H-K)_{\rm MKO} + (0.54 \pm 0.04)$.
Compared to the original CTTS locus, $(J-H)_{\rm CIT} = (0.58 \pm 0.11)
\times (H-K)_{\rm CIT} + (0.52 \pm 0.06)$, the MKO locus has smaller
$J-H$ with increasing $H-K$. This tendency is consistent with that for
the dwarf track as discussed above.

%
Lastly, we considered the reddening vector, which is the 
the slope $(J-H)/(H-K)$ and the length of the $A_V = 5$ mag vector.
%
Using the near-infrared camera SIRIUS equipped with the MKO filters,
\citet{IRSF} estimated that $(J-H) / (H - K_S) = 1.72$ from the data
toward the Galactic Center. This value is almost identical to the
conventional slope ($J-H / H-K = 1.70$) in Arizona system \citep{RL}.
Although IRSF uses a $K_S$ band filter instead of a $K$ band filter, the
difference of the filter profile is so small \citep[see Fig.~1
in][]{Tokunaga2002} that the magnitude differences are
negligible\footnote{For reference, we compared $K$ and $K_S$ magnitudes
of Persson standard stars \citep{Persson1998} and found that the
magnitude differences are almost within 0.02 mag (the average difference
is $-0.0070 \pm 0.0067$ mag). Although these magnitudes are in the LCO
photometric system \citep{Persson1998}, similar results would be
expected for the MKO photometric system because the difference of the
MKO $K_S$ and $K$ filter profiles is smaller than that of LCO $K_S$ and
$K$ filter profiles.}.  Therefore, we could conclude that the reddening
vector is almost identical for both Arizona and MKO systems.  In the
following section, we use the MKO color-color diagram for all the
analysis.


\section{Color-Color Diagram for Low Metallicity Environment}
\label{sec:LowCol}

Because the metallicity in EOG is significantly lower than in the solar
neighborhood, we should consider the effect of low metallicity on the
color-color diagram.
Following the PMS models by \citet{{D'Antona1997},{D'Antona1998}}, we
suggested that NIR absolute magnitudes of PMS stars with $- 0.3$ dex
(1/2 solar metallicity) is not significantly different from those with
the solar metallicity \citepalias{My2006ApJ}. Therefore colors also
should not significantly change down to a metallicity of $- 0.3$ dex.
In this section we discuss the effect of low metallicity extended to
$-1$ dex on the $J-H$ and $H-K$ colors considering both observations and
models in the literature.

\citet{Leggett1992} presented NIR photometry of low-mass stars in young
disk ([M/H] $\sim 0.0$), young-old disk, old disk ([M/H] $\sim -0.5$),
old disk-halo ([M/H] $\sim -1.0$), and halo ([M/H] $< -1.0$), showing
that $J-H$ colors of dwarfs between K7 (0.5 $M_\odot$) and M5 ($\sim
0.1$ $M_\odot$) types
decrease by $\sim 0.1$ mag with decreasing metallicity
(Fig.~\ref{fig:LeggettLowMeta}).
Similar tendency is seen in the model by \citet{Girardi2002}, who listed
the NIR magnitudes of stars with mass of 0.15--1 $M_\odot$ for various
metallicities ($Z= 0.001, 0.004, 0.008,$ and 0.019).
%
%
Although the color values are slightly different for the observation and
the Girardi model, both show the quite similar color variation 
$\Delta (J-H) = -0.1$ for $\Delta$ [M/H] $\sim -1$,  which strengthens
the validity of the metallicity dependence.
However, for the metallicity of Cloud 2 ([M/H] $\sim -0.5$), the dwarf
track is still almost identical to the standard B \& B dwarf track
within the uncertainty (Fig.~\ref{fig:LeggettLowMeta}).
As for the stars with mass of $\ga 1$ $M_\odot$, \citet{Girardi2002}
suggested that the colors weakly depend on metallicity.

%
\citet{Nakajima}, who observed the LMC ($-0.5$ dex) in the MKO system,
suggested that the slope of $(J-H) / (H-K)$ for LMC appears to be
consistent with that for solar metallicity \citep{IRSF}, suggesting that
the reddening vector for the LMC metallicity is not significantly
different from that for the solar metallicity.

We conclude that dwarf track and reddening vector in $JHK$ color-color
diagrams do not strongly depend on the metallicity down to that of the
stars in Cloud 2 ($\sim -0.7$ dex).
Therefore, we do not include the effects of low metallicity in the
following analysis using color-color diagram, such as the identification
of cluster members ($\S$~\ref{sec:IDENT}), the estimation of extinction
and color excess ($\S$~\ref{sec:AvHK0}).


\section{RESULTS}
\label{sec:RESULTS}

\subsection{Identification of Cluster Members} \label{sec:IDENT}

\begin{figure}
\epsscale{1.1}
\includegraphics[scale=2]{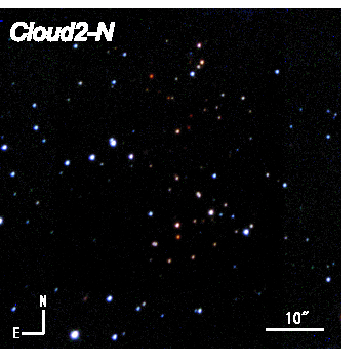}
\hspace{1cm}
\includegraphics[scale=.538]{f1R.eps}
%
\caption{{\it Left}: $JHK$ pseudocolor image of the Cloud 2-N
cluster. North is up and east is left.  The field of view is about $1'
\times 1'$.  The coordinate of the field center is ($\alpha_{\rm 2000},
\delta_{\rm 2000}$) = ($02^h 48^m 42.5^s, +58^\circ 28' 56\farcs0$) with
an uncertainty of less than 0.5 arcseconds.
{\it Right}: Spatial distribution of Cloud 2-N cluster members. Filled
circles and open circles represent cluster members and field stars,
respectively.}  \label{fig:3colN}
\end{figure}

\begin{figure}
\includegraphics[scale=2]{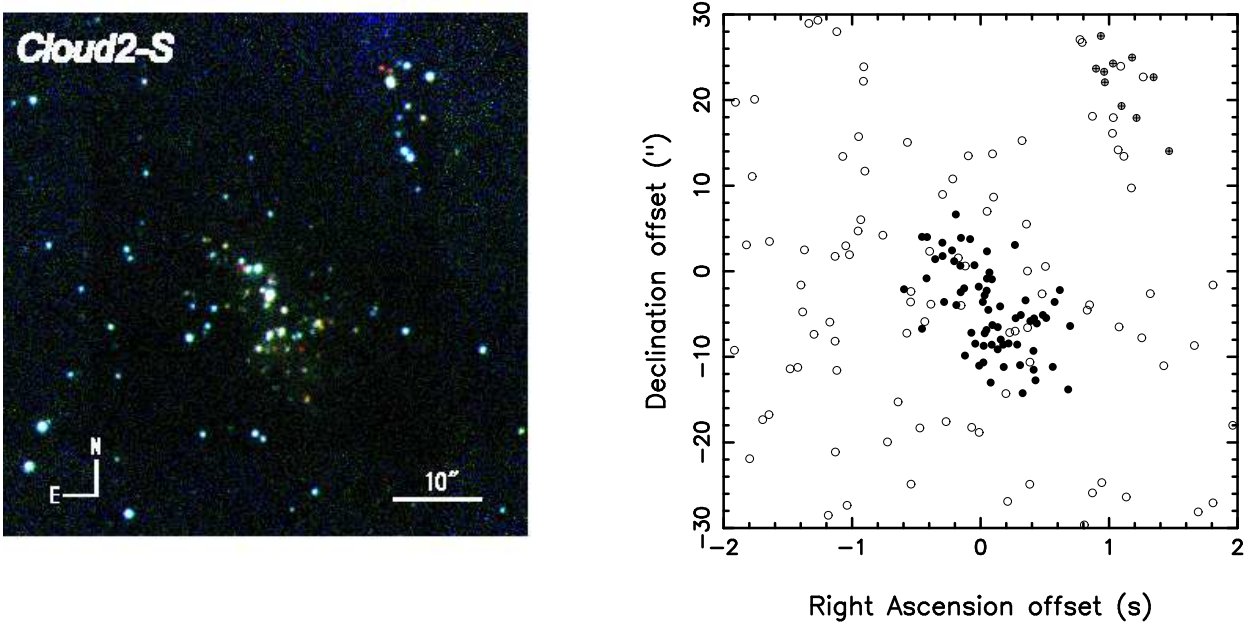}
\hspace{1cm}
\includegraphics[scale=.538]{f2R.eps}
%
\caption{{\it Left}: $JHK$ pseudocolor image of the Cloud 2-S
cluster. North is to up and east is left. The field of view is about $1'
\times 1'$.  The coordinate of the field center is ($\alpha_{2000},
\delta_{\rm 2000}$) = ($02^h 48^m 28.7^s, +58^\circ 23' 34\farcs6$) with
an uncertainty of less than 0.5 arcseconds.
{\it Right}: Spatial distribution of the members of the Cloud 2-S
cluster, the Cloud 2-S sub-cluster and field stars are shown with filled
circles, circles with cross, and open circles, respectively.  The Cloud
2-S sub-cluster was designated at IRS3 in \citet{KT2000} as discussed in
$\S$~\ref{sec:IDENT}.}
\label{fig:3colS}
\end{figure}

We conducted identification of the cluster members using pseudocolor
pictures of the observed fields (Fig.~\ref{fig:3colN}, \ref{fig:3colS})
and $(J-H)$ vs. $(H-K)$ color-color diagrams of all the detected sources
(Fig.~\ref{fig:colcolNS}).
We identified cluster members which have color of $(H-K) \ga 0.5$ and are
within the cluster regions as identified in Kobayashi et al. (2007).
Because there are no foreground molecular clouds in front of Cloud 2
(Kobayashi \& Tokunaga 2000), the red colors of the cluster members
should originate from the extinction by Cloud 2 itself as well as by
circumstellar material (see $\S$~5.2 for more discussion). Contamination
from the background sources should be very small in view of the large
$R_g$ of Cloud 2.

We found 72 cluster members in the Cloud 2-N cluster and 66 cluster
members in the Cloud 2-S cluster.  We also identified 10 YSOs in the
Cloud 2-S sub-cluster (Kobayashi et al. 2007), which is $\sim 40''$
northern-east from the Cloud 2-S cluster. Fig.~\ref{fig:3colN} and
Fig.~\ref{fig:3colS} show the spatial distributions of these cluster
members and field stars in the Cloud 2-N and -S fields, respectively.
Because of the re-analysis with the fainter detection limit, the number
of identified Cloud 2-N cluster members has increased from 52 in
\citetalias{My2006ApJ} to 72.

\subsection{Extinction and Disk Color Excess of Cluster Members}
\label{sec:AvHK0}

The extinction and disk color excess for each star were estimated using
the color-color diagram.
For reliable estimation of the parameters, only stars in the positions
reddened from the CTTS locus and the dwarf locus are used (42 cluster
members and 36 field stars for the Cloud 2-N cluster, 45 cluster members
and 45 field stars for the Cloud 2-S cluster). For convenience the dwarf
locus was approximated by the extension of the CTTS locus drawn out to
$H-K \sim 0.1$ mag. In the color-color diagram the extinction $A_V$ of
each star was estimated from the distance along the reddening vector
between its location and the stellar loci. The resultant $A_V$
distributions of Cloud 2-N and Cloud 2-S cluster members
(Fig.~\ref{fig:AvNS}) have peak at 7.2 and 6.1 mag, respectively ($A_K =
0.81$ and 0.68 mag) and those for field stars in the both Cloud 2 fields
have a peak at 2.2 mag ($A_K = 0.25$ mag). The distributions of the both
clusters' members show that the both clusters are reddened uniformly by
about $A_V \sim 4$--5 mag inside Cloud 2. This verifies the selection
method of cloud members as described in $\S$~\ref{sec:IDENT}.
These values of $A_V$ are also consistent with the estimated from sub
radio and continuum observation (Ruffle et al. 2007).

We also constructed the distributions of the unreddened color $(H-K)_0$
of the cluster members and the field stars (Fig.~\ref{fig:HKNS}).
For both clusters, the distributions of the cloud members and field
stars show the clear peak offset of about 0.1 mag.
The difference of the {\it average} $(H-K)_0$ between the cluster
members and the field stars (Cloud 2-N cluster: 0.26 mag, Cloud 2-S
cluster: 0.37 mag) can be attributed to thermal emission of
circumstellar disks of the cluster members. Assuming that disk emission
appears in the $K$~band but not in the $H$~band,
the disk color excess of the Cloud 2-N and -S cluster members in the
$K$~band, $\Delta K_{\rm disk}$, is equal to 0.26 and 0.37 mag,
respectively.


\section{DISCUSSION}
\label{sec:DIS}

\subsection{Stellar Density Variation}
\label{subsec:SD}

\citet{AnnualReview} compiled a catalog of embedded clusters that are
located in $D \lesssim 2$ kpc (distance modulus DM $< 11.5$ mag) and
detected cloud members with the apparent limiting magnitudes of $m_K
\sim 16.0$ mag. \citet{Carpenter2000_2mass} analyzed 2MASS data of the
Perseus, Orion A, Orion B, and Monoceros R2 clusters at $D \simeq 500$
pc (DM $\simeq 8.5$ mag) and detected members with apparent limiting
magnitudes of $m_{K_S} = 14.3$ mag. Both catalogs thus consist of
clusters that are detected with absolute limiting magnitude of $M_K \sim
5$--6 mag in the solar neighborhood. In the present study, since the
Cloud 2 clusters at $D \sim 12$ kpc (DM $=15.4$ mag) are detected with
apparent magnitudes of $m_K \sim 20.5$ mag, the absolute limiting
magnitudes is $M_K \sim 5$ mag. Therefore, we can compare the stellar
density of embedded clusters in EOG to those of embedded clusters in the
solar neighborhood on the same basis for the first time.

\citet{Adams06} found a correlation between the number of stars in a
cluster and the radius of the cluster, using the tabulated cluster
properties in \citet{AnnualReview} and \citet{Carpenter2000_2mass}
(Fig.~\ref{fig:SteDen}).  \citet{Allen07} suggested that the average
stellar density of clusters have a few to a thousand members varies by a
factor of only a few.  Stellar densities of the Cloud 2-N and -S
clusters were estimated at 13 pc$^{-2}$ and 50 pc$^{-2}$ from the
spatial distribution of the identified cluster members in the $2 \times
2$ pc$^2$ area and a circle of 0.6 pc radius, respectively.  For the
Cloud 2-N cluster, the above density is a little larger than our old
estimate in \citetalias{My2006ApJ} ($\sim 10$ pc$^{-2}$) because we
included the newly identified fainter stars.  In Fig.~\ref{fig:SteDen},
the Cloud 2-N cluster appears to be a kind of ``loose cluster'', while
the Cloud 2-S cluster appears to be a kind of ``dense cluster''.  For
the typical stellar density of the clusters ($\sim 50$ pc$^{-2}$, see
Figure~\ref{fig:SteDen}), the average projected distance between two
adjacent stars is $\sim 0.14$ pc, which corresponds to $\sim 2''$ at
$D=12$ kpc.  For even densest clusters with $\sim 1000$ pc$^{-2}$, the
average projected distance is $\sim 0.03$ pc, or $0\farcs5$ at $D=12$
pc. These separations are sufficiently resolved for this observation.
Even if all the stars are unresolved binaries, the ``loose'' and
``dense'' characteristics of the Cloud 2 clusters do not change (see
gray circles in Figure~\ref{fig:SteDen}).
The difference of the stellar densities is very interesting, because the
two clusters show very similar characteristics for other points as shown
in the following subsections.  Kobayashi et al. (2007) attributed the
difference of the star formation mode to the ambient pressure difference
of the two clusters.

\subsection{IMF and Age of the Clusters}
\label{subsec:KLF}

$K$-band luminosity functions (KLFs), which are the number of stars as a
function of $K$-band magnitude, of different ages are known to have
different peak magnitudes and slopes (Muench et al. 2000).
IMFs and ages of young clusters can be estimated by comparison of the
observed KLF with model KLFs, which are constructed from IMFs and
$M$-$L$ relations.
In this section, we use the model KLF from \citetalias{My2006ApJ} (see
$\S$~4).
Although the popular PMS evolution models such as
\citet{{D'Antona1997},{D'Antona1998}}, \citet{Siess2000}, and
\citet{BCAH97}, do not have models for metallicity less than 1/2 solar
($-0.3$ dex).
We assumed that the $M$-$L$ relation for solar metallicity can be
applied to model KLFs for low metallicity because the difference of
$M$-$L$ relations for the solar metallicity and those for 1/2 solar
metallicity is very little (see more detail in Paper I).

Following the method in \citetalias{My2006ApJ}, we constructed KLFs of
the Cloud 2 cluster members (Fig.~\ref{fig:KLFNSandCOMP}).
We estimated the detection completeness in each magnitude bin by putting
artificial stars on random positions in the field and checking whether
they are detected in the same way as for the real objects.
Five stars are placed at a time in each magnitude bin and the check was
conducted 200 times, resulting in about 1000 artificial stars per
magnitude bin. The result is also shown in Fig.~\ref{fig:KLFNSandCOMP}
(bottom), and the completeness-corrected KLFs are shown in
Fig.~\ref{fig:KLFNSandCOMP} (top) with thick lines.
%
Both KLFs 
appear to be very similar in the slope at brighter magnitude side and
the peak position ($\sim 19$ mag).

The comparison of the completeness corrected KLFs with the model KLFs
are shown in Fig.~\ref{fig:KLFmodel}. Because the slope of model KLF
does not significantly depend on small difference of underlying IMFs
\citepalias[see Fig.~6, 7 in][]{My2006ApJ}, we used the Trapezium
IMF\citep{Muench2002}, which is thought to be the most reliable IMF for
young clusters (e.g., Lada \& Lada 2003).
A quick visual check of Fig.~\ref{fig:KLFmodel} suggests that the ages
of the Cloud 2-N and -S clusters are similar and it is $\sim 0.5$--1
Myr.
It is difficult to estimate the age of the cluster with an accuracy of
0.1 Myr because isochrone models for ages of less than 1 Myr is thought
to be uncertain \citep[e.g.,][]{Baraffe2002}.
However, we can at least conclude that the ages of the two clusters are
no more than 1 Myr \citepalias[see $\S$~5.1 in][]{My2006ApJ}.
For the Cloud 2-N cluster, twenty cluster members are detected in the 
$K \sim 20$ magnitude bin and more in the fainter magnitude bins, while
only four members were detected in the $K \sim 20$ magnitude in (the
faintest bin) \citetalias{My2006ApJ}.
However, this change in the fainter bins ($K \ga 20$) did not affect the
age estimation, because the brighter bins ($K < 19$) are more sensitive
for the KLF fitting.
The relatively good fit of the model KLF to the observed KLF suggests
that the IMF at $M > 0.1$ $M_\odot$ in Cloud 2 is not significantly
different from the ``universal'' IMF for Trapezium (see also discussion
in Paper I).

The similarities of the KLFs and the estimated ages for the two
independent clusters with a large separation ($\sim 25$ pc) strongly
suggest that their star formation activities were triggered by the same
mechanism.
The ages of the clusters were estimated at $\sim 0.5$--1 Myr, which is
much less than that of the SNR \ion{H}{1} shell (GSH 138-01-94; Stil \&
Irwin 2001), 4.3 Myr.
An independent age estimate of the Cloud 2-N cluster ($\sim 0.5$--1 Myr)
by Kobayashi et al. (2007) is in quite good agreement with our estimate.
These age estimates strongly support the idea that the star formation
activity in Cloud 2 was triggered by the SNR \ion{H}{1} shell, as
discussed in \citetalias{My2006ApJ} and Kobayashi et al. (2007).


\subsection{Star Formation Efficiency (SFE)}

The star formation efficiency (SFE), which is generally defined as [SFE
= $M_{\rm stars} / (M_{\rm gas} + M_{\rm stars})$] where $M_{\rm gas}$
is cluster-forming core mass and $M_{\rm stars}$ is the total stellar
mass, is one of the most fundamental parameters of the star and
cluster-formation processes \citep{AnnualReview}.  The estimated SFEs of
the Cloud 2 clusters could be ``pure'' SFE of one-time cluster formation
event because there has been no complex star formation history in the
EOG.  As discussed in $\S$~\ref{subsec:SD}, the detection of the
embedded cluster members in the EOG with the high-sensitivity ($m_K
\simeq 21.0$ mag) enables a fair comparison of SFE in the EOG with that
in the solar neighborhood.  However, because SFEs have been estimated
\citep[e.g.,][]{{Lada92L},{Lada97}, {Hartmann02}} in various ways,
ex. $M_{\rm gas}$ is estimated using different kinds of molecular
emission lines ($^{12}$CO, $^{13}$CO, and C$^{18}$O), $M_{\rm stars}$ is
estimated with a very simple assumption that all stars have a mass of 1
$M_\odot$, 0.5 $M_\odot$, etc., here we attempted to estimate SFEs of
various star forming regions in a uniform fashion as described in the
following paragraphs.

For the estimates of $M_{\rm gas}$ we use $^{13}$CO because of the
relatively high sensitivity even for distant molecular clouds and also
the wide availability of the archival data.
Because $M_{\rm gas}$ from $^{13}$CO correlates well with those from
C$^{18}$O \citep[C$^{18}$O--$^{13}$CO mass correlation;][Fig.~18
(a)]{Ridge2003}, it is possible to use C$^{18}$O data when $^{13}$CO
data is unavailable.
For the accurate estimation of $M_{\rm stars}$, it is necessary to take
the mass of the low mass stars into account.
We use a summary of $M_{\rm stars}$ for star forming molecular clouds in
the solar neighborhood \citep[$D \la 2$ kpc;][Tab.~1]{AnnualReview}.
They derived $M_{\rm stars}$ for each cluster by predicting infrared
source counts, down to 0.017 $M_\odot$ stars, as a function of differing
limiting magnitudes using the KLF models of \citet{Muench2002}.
As an example, we estimated the SFE of the well-known embedded cluster,
NGC 1333, which is in the star forming region in Taurus, at 5.2 \% using
$M_{\rm gas} = 1450$ $M_\odot$ from $^{13}$CO (Warin et al. 1996) and
$M_{\rm stars} = 79$ $M_\odot$.
We also estimated the SFE of the
NGC 2024 at 17 \% using $M_{\rm gas} = 896$ $M_\odot$ from C$^{18}$O
(Lada et al. 1991a,b) and the C$^{18}$O--$^{13}$CO mass correlation
(Ridge et al. 2003), and $M_{\rm stars} = 182$ $M_\odot$. 
Of 76 embedded clusters in the solar neighborhood \citep{AnnualReview},
the SFEs for 30 embedded clusters, whose $M_{\rm gas}$ could be found in
the literatures, are estimated in the same way, resulting in the SFEs
widely ranging from 2.3 to 57 \% with an median of 15 \%.  Median size
and mass of the cluster-forming cores are $0.6$ pc (0.32--1.0 pc in
FWHM) and $M_{\rm gas} = 480$ $M_\odot$ (27--4000 $M_\odot$),
respectively, while median size and mass of embedded clusters are 0.62
pc (0.3--1.9 pc) and $M_{\rm stars} = 66$ $M_\odot$ (25--340 $M_\odot$),
respectively.

The spatial distribution of the Cloud 2 cluster members (red crosses)
and the molecular cloud cores (blue contour for $^{13}$CO and gray scale
for $^{12}$CO) are shown in Fig.~\ref{fig:CO_NIR}.
$M_{\rm stars}$ of the Cloud 2-N and -S clusters are estimated at 48 and
46 $M_\odot$, respectively, 
by directly counting the stellar mass of all cluster members larger than
mass detection limit ($< 0.1 M_\odot$) using the model KLF at the age of
0.5 Myr constructed in $\S$~\ref{subsec:KLF}.
Even if we estimate $M_{\rm stars}$ with the stars down to 0.017
$M_\odot$ as the same way in Lada \& Lada (2003), the $M_{\rm stars}$ 
increases only $\sim 5$ \% .
Both sizes of the clusters are typical, whose radii are $\sim 1$ pc for
Cloud 2-N and $\sim 0.6$ pc for Cloud 2-S.
For $M_{\rm gas}$, because of the large distance to Cloud 2 and the
relatively low spatial resolution of radio observations compared to NIR
observations,
the region of associated core must be carefully assessed.  In
Fig.~\ref{fig:CO_NIR} the Cloud 2-N core has two sub-peaks which have
similar flux, while the Cloud 2-S core appears to be single.
Because the Cloud 2-N cluster is associated with the northern sub-peak,
we have estimated the $M_{\rm gas}$ for the Cloud 2-N cluster using only
the northern part of the cloud.
%
%
The resultant $M_{\rm gas}$ for the Cloud 2-N and Cloud 2-S clusters are 360
$M_\odot$ and 310 $M_\odot$ with the radii of 2.7 and 1.7 pc in FWHM,
respectively (Saito et al. 2007).
As a result, the SFEs of Cloud 2-N, -S clusters are estimated at 12 and
13 \%, respectively. These SFEs are comparable to embedded clusters in
the solar neighborhood.
For FOG, there have been several studies on SFE of star forming
molecular clouds. 
%
%
\citet{Snell2002} found that the ratio of far-infrared luminosity to
molecular cloud mass in FOG is comparable to that of molecular clouds in
the solar neighborhood, suggesting SFE in FOG is similar to that in the
solar neighborhood.
Our estimates of SFEs in EOG are consistent with these results for FOG.
It is also interesting to note that the SFEs for the two independent
clusters are similar despite the large difference in the stellar
densities (see $\S$~\ref{subsec:SD}).

Note that the estimated SFEs still have some systematic uncertainties
that should be carefully considered.
Firstly, it is possible that some stars were not counted because of
confusion with other stars.  For example we might miss more close
binaries at this large distance.  However, even if all stars are binary
with the same mass, the SFEs of the Cloud 2-N and -S clusters are 24
\% and 26 \% respectively. This is not unusual.
Secondly, it is possible that we are not still resolving the cloud cores
 because of the large distance to Cloud 2, thus overestimating the cloud
 mass by including the mass of unrelated cores.  Actually, the radii of
 Cloud 2 cores ($> 2$ pc) are about twice  the maximum radius of
 nearby cores ($< 1$ pc), there is a possibility that the Cloud 2 cores
 are not resolved by a factor of two.  In this case, SFEs are
 underestimated.
Lastly, $M_{\rm gas}$ may be underestimated given low metallicity of
Cloud 2 because we assume the normal fractional abundance of
$X(^{13}$CO) at $2 \times 10^{-6}$ in deriving $M_{\rm gas}$. Since the
metallicity is lower by a factor of 5 than that in the solar
neighborhood, the $X(^{13}$CO) is likely 5 times larger than the
assumed value. As a result, $M_{\rm gas}$ may become larger accordingly.
%
%
In this case, SFEs are overestimated.  Of these uncertainties the
estimate of $M_{\rm gas}$ is most critical.  For more accurate estimate
of SFEs, radio observation with higher spatial resolution (e.g., with
interferometry) is necessary and a more accurate CO-H$_2$ conversion
factor $X(^{13}$CO) as a function of metallicity is needed.

\section{CONCLUSION}

We have conducted a pilot study of star formation in EOG using the two
embedded clusters in Cloud 2. Our conclusions can be summarized as
follows:

\begin{enumerate}
%
\item We have obtained deep near-infrared images of two embedded
      clusters in EOG, the Cloud 2 clusters, with Subaru 8.2 m telescope
      and the IRCS infrared imager. This is the first deep infrared
      imaging of star forming regions in EOG, with the mass detection
      limit close to the substellar regime ($< 0.1 M_\odot$).

%
\item Due to the faint detection-limit, we investigated the IMF and age
      of the clusters in EOG by identifying the peak magnitude of the
      KLF. The underlying IMF of the cluster down to the detection limit
      is not significantly different from the typical IMFs in the field
      and in the nearby star clusters.  For investigating the behavior
      of IMF (KLF) in the substellar regime, even deeper imaging is
      required and we might find a different IMF than the solar
      neighborhood.  We have estimated the age of the clusters to be
      $\sim 0.5$--1 Myr.
      Additionally, SFEs of the Cloud 2 clusters were comparable ($\sim
      10$ \%) within the uncertainties to nearby embedded clusters.
      Despite these identical characteristics of the Cloud 2 clusters,
      the estimated stellar density suggests that different types of
      star formation are ongoing: an ``isolated type'' star formation is
      ongoing in Cloud 2-N, while a ``cluster type'' star formation is
      ongoing in Cloud 2-S.

\item We summarized the existing models of near-infrared characteristics
      of low metallicity stars to derive the mass-luminosity relation
      and color-color diagram for low metallicity stars.
      We concluded that dwarf track and reddening vector in $JHK$
      color-color diagrams do not strongly depend on the metallicity
      down to that of the stars in Cloud 2 ($\sim -0.7$ dex).
      Therefore, we did not include the effects of low metallicity in the
      following analysis using color-color diagram, such as the identification
      of cluster members,
      the estimation of extinction and color excess. 
      However, the effect could be significant for even lower
      metallicity and more theoretical studies and compilations of the
      near-infrared data for low-metallicity normal stars are needed.

%
\item For interpreting the photometric result, we have constructed the
      NIR color-color diagram (dwarf track, CTTS locus, reddening
      vector) in the MKO filter system, for which colors for main
      sequence stars have not been well established before. We find that
      it is important to carefully compare colors in different
      photometric systems.
      
\end{enumerate}

\acknowledgments

We are grateful to Yosuke Minowa at NAO Japan for kindly introducing us
the use of SExtractor and giving us fruitful comments for completeness
correction.  The data presented here was obtained during the commission
phase of Subaru Telescope and IRCS. We truly thank all the Subaru staff
who made these observations possible.
This research was supported by Grant-in-Aid for Encouragement of
Scientists(A) of the Ministry of Education, Science, Culture, and Sports
in Japan (No.11740128).
%
%
%
C.Y. is financially supported by a Research Fellowship from the Japan
Society for the Promotion of Science for Young Scientists. 








\clearpage

%
%


\begin{figure}[htpb]
\begin{center}
\epsscale{1.1}
\plottwo{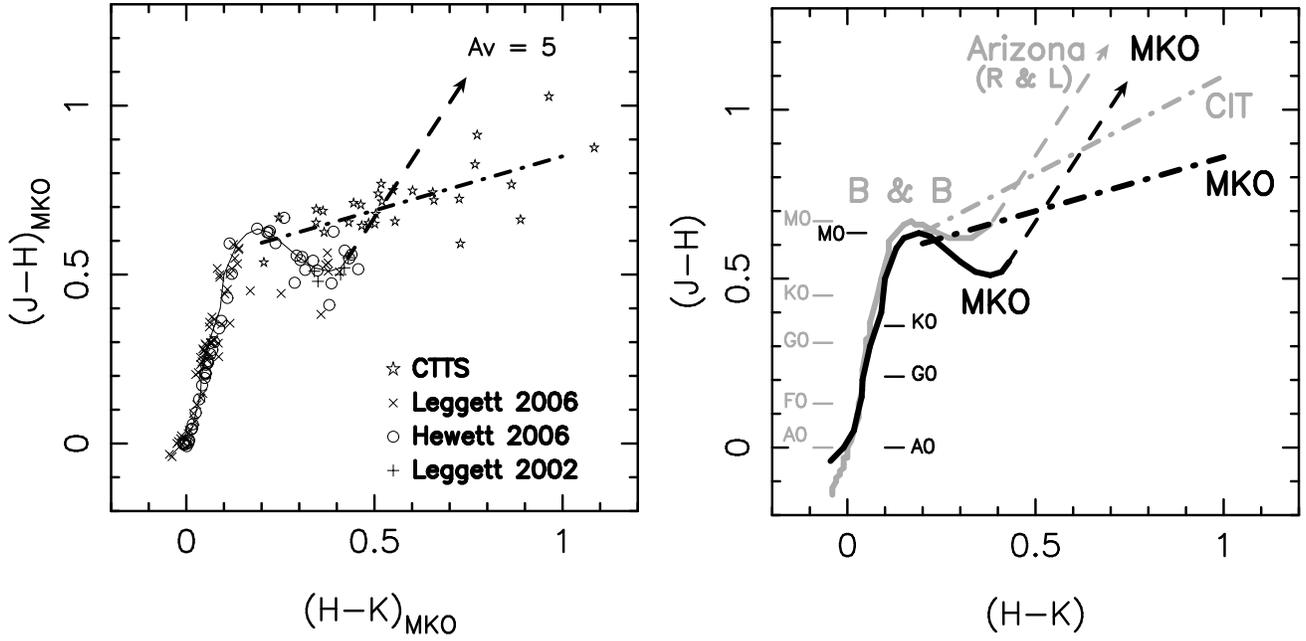}{f3b.eps}
%
\caption{{\it Left}: ($J-H$) vs. ($H-K$) diagram in the MKO photometric
system.  The dwarf star track (solid line) is constructed from colors of
dwarf stars observed with MKO filters (Hewett et al. 2006, open circle,
B9V--M6; Leggett et al. 2002, plus symbol, M4V--M6V; Leggett et
al. 2006, cross symbol, M4--M6). See the details in the main text. CTTS
locus (dot-dashed line) is estimated by least-square fit of the CTTS
data (star symbol). The reddening vector for $A_V = 5$ mag is shown with
a dashed arrow from M6V location on the dwarf star track.
{\it Right}: Comparison of dwarf star tracks (solid lines), CTTS loci
(dot-dashed lines), and reddening vectors (dashed arrows) in MKO (black)
and other photometric systems (gray).  For the latter, the dwarf star
track is in Johnson-Glass system \citep{Bessell}, the reddening vector
is in Arizona system \citep{RL}, and the CTTS locus is in CIT system
\citep{T Tauri}.}  \label{fig:colMKO}
\end{center}
\end{figure}


\begin{figure}[htpb]
\begin{center}
\epsscale{0.6}
\plotone{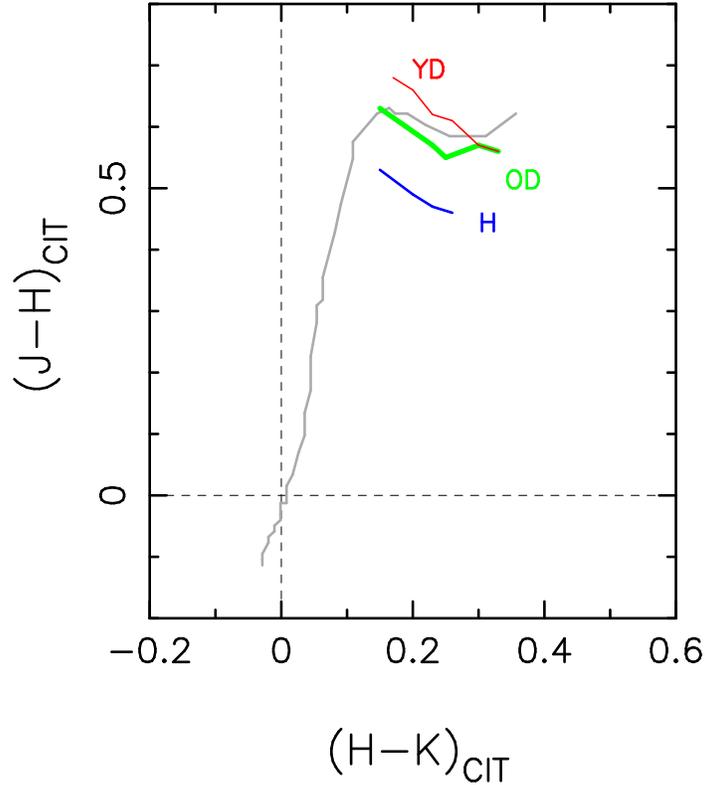}
%
\caption{Dwarf star tracks of differing metallicities in $JHK$
color-color diagram. Note that all the colors are shown in CIT system.
Color lines are low-mass star tracks from Table~6 in
\citet{Leggett1992}.  YD ([M/H] $\sim 0.0$; M0--M6), OD ([M/H] $\sim
-0.5$; M0--M5.5), and H ([M/H] $< -1.0$; K7--M3) represent stars in
young disk, old disk, and halo, respectively.
The gray line shows the standard dwarf track from \citet{Bessell} after
converting the color system from Johnson-Glass to CIT using the
transformation in \citet{Leggett1992}.}  \label{fig:LeggettLowMeta}
\end{center}
\end{figure}





\begin{figure}
\epsscale{0.45}
\plotone{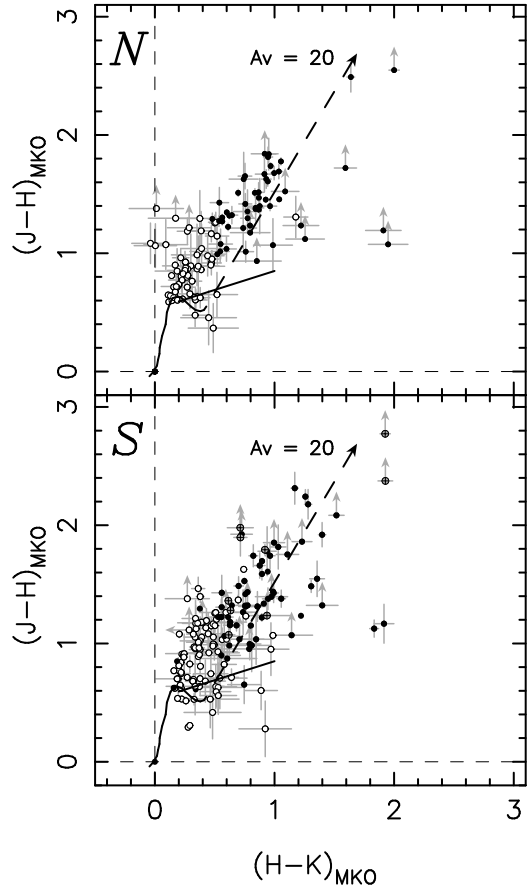}
%
\caption{($J-H$) vs. ($H-K$) color-color diagrams of the Cloud 2-N
cluster (top) and the Cloud 2-S cluster (bottom). Identified cluster
members and field stars are shown with filled circles and open circles,
respectively.  Cloud 2-S sub-cluster members are shown with circled
crosses (bottom).  Only stars detected with more than $5\sigma$ in all
$JHK$-bands are plotted. The dwarf star track, CTTS locus, and reddening
vector are based on the MKO photometric system (see
Fig.~\ref{fig:colMKO} and discussion in $\S$~\ref{sec:colMKO},
$\S$~\ref{sec:LowCol}).  For stars that are not detected with 5 $\sigma$
in $J$-band, we showed the $J-H$ upper limit.} \label{fig:colcolNS}
\end{figure}




\begin{figure}
\epsscale{1.1}
\plottwo{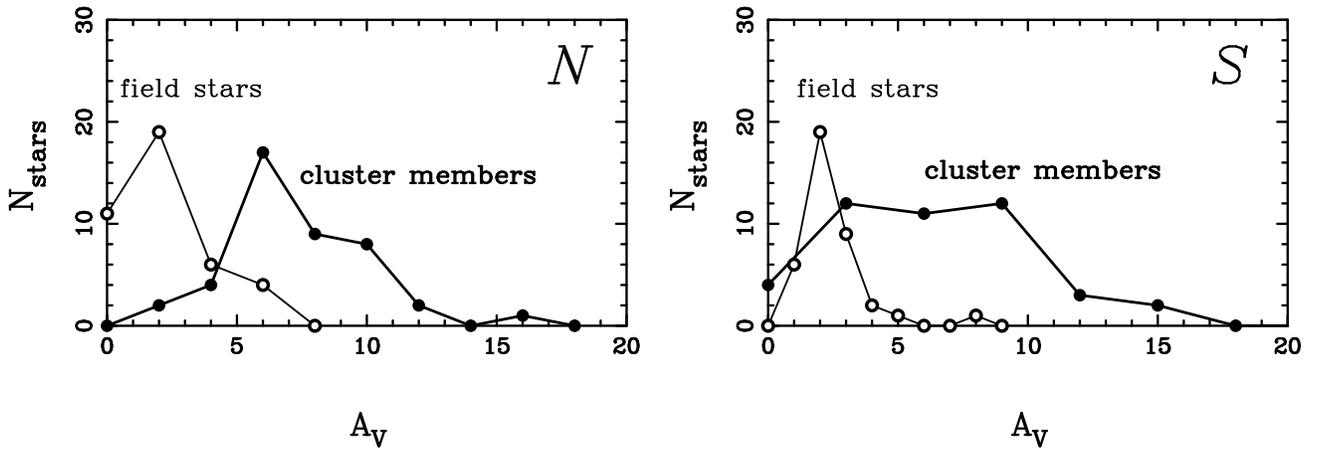}{f6b.eps}
%
\caption{$A_V$ distribution of Cloud 2 cluster members (thick lines) and
field stars (thin lines). Average $A_V$ values of the members in Cloud
2-N and -S clusters are 7.2 and 6.1 mag, respectively, while those field
stars in the Cloud 2-N and -S clusters are both 2.2 mag, respectively.
Note that Cloud 2-S sub-cluster members are not counted in the right
plot.}  \label{fig:AvNS}
\end{figure}

%

\begin{figure}
\epsscale{1.1}
\plottwo{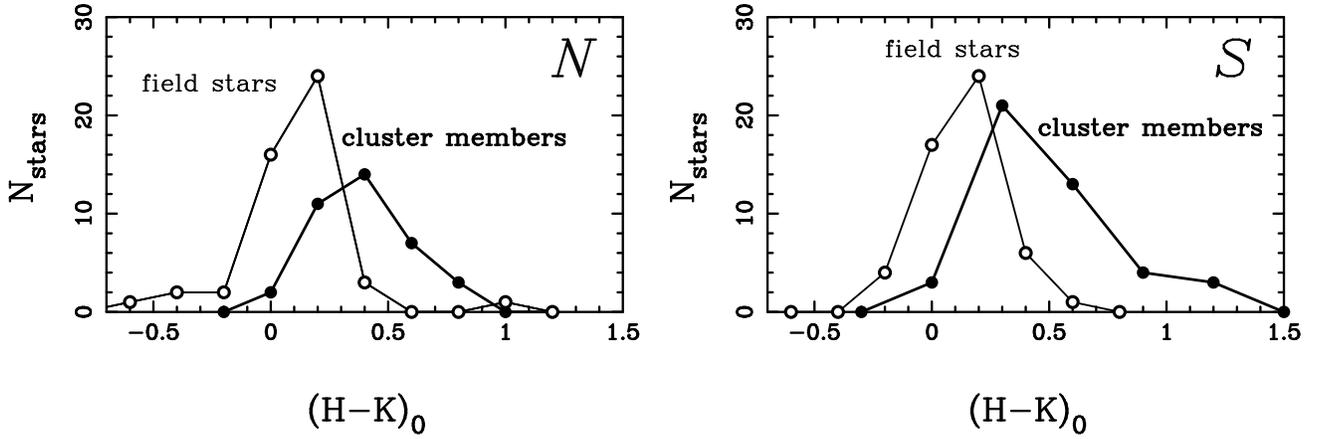}{f7b.eps}
%
%
%
\caption{$(H-K)_0$ distributions of Cloud 2 clusters' members (thick
lines) and field stars (thin lines).  Average $(H-K)_0$ values of the
cluster members in Cloud 2-N and Cloud 2-S clusters are 0.37 and 0.49 mag,
respectively, while those field stars in Cloud 2-N and -S clusters are
0.11 and 0.12 mag, respectively.} \label{fig:HKNS}
\end{figure}




\begin{figure}
\epsscale{0.65}
\plotone{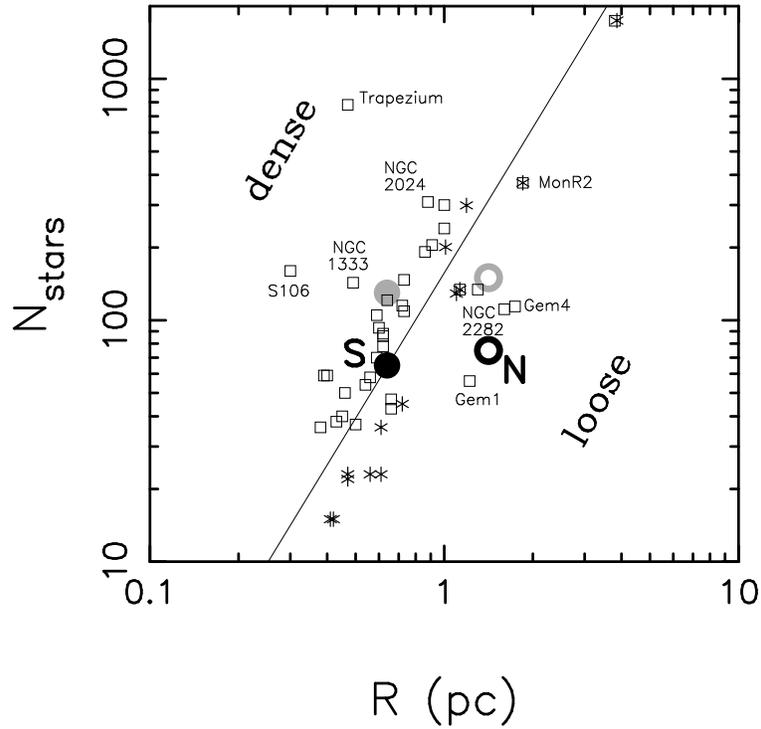}
%
%
\caption{$N_{\rm star}$ vs. cluster radius for the Cloud 2-N cluster
(open circle), Cloud 2-S cluster (filled circle), plotted along with the
data for nearby embedded clusters from \citet[][squares]{AnnualReview}
and Carpenter (2000, stars).
The solid line shows the constant stellar density of 50 pc$^{-2}$. The
gray open and filled circles show the points for the Cloud 2 clusters
for the case that all the stars are unresolved binaries.}
\label{fig:SteDen}

\end{figure}


\begin{figure}
\epsscale{0.6}
\plotone{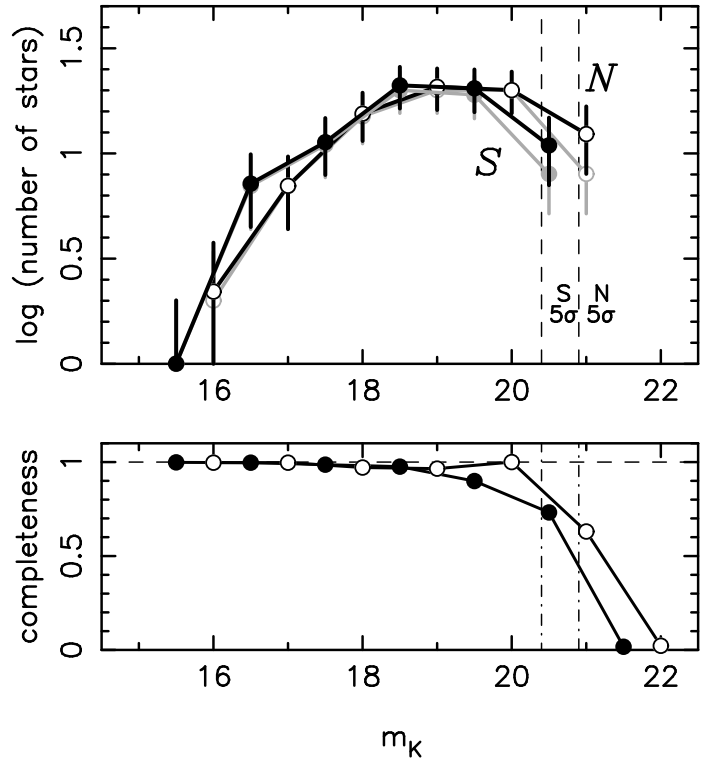}
%
%
\caption{ {\it Top}: The raw KLF (gray lines) and completeness-corrected
KLF (black lines) of Cloud 2 clusters. The observed KLFs for the Cloud
2-N and Cloud 2-S clusters are shown with filled circle and open circle,
respectively.  Error bars show the uncertainties due to Poisson
statistics. Dashed lines show the limiting magnitudes of 5 $\sigma$
detection (Cloud 2-N: 20.9 mag, Cloud 2-S: 20.4 mag).
{\it Bottom}: Detection completeness curves derived from the simulations
using the artificial point sources. See the details in the main text.}
\label{fig:KLFNSandCOMP}
\end{figure}

\begin{figure}
\epsscale{1.1}
\plottwo{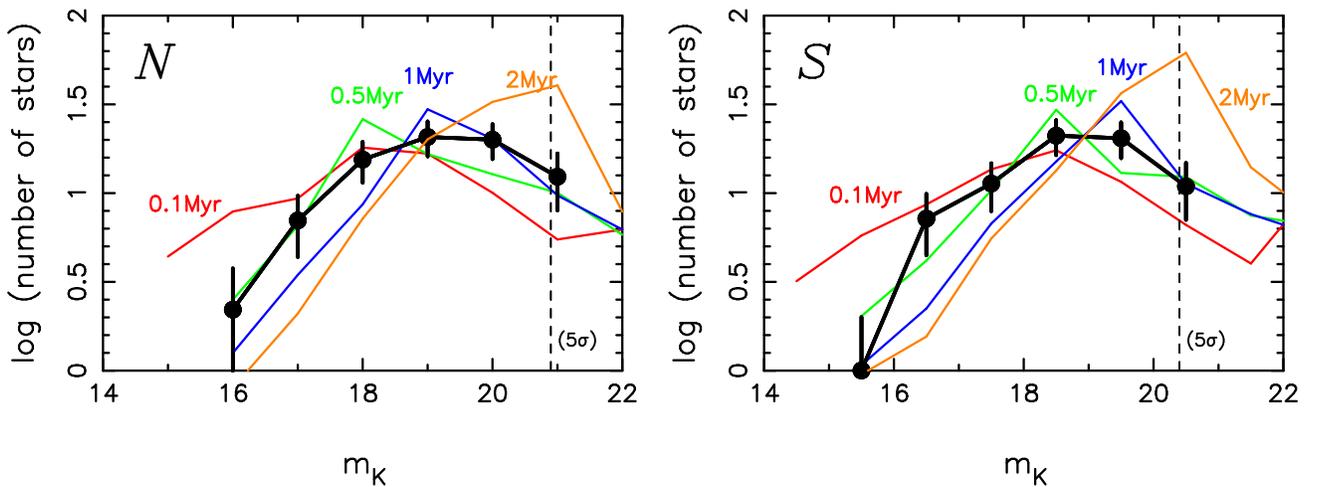}{f10b.eps}
%
%
\caption{Comparison of completeness-corrected KLF (thick line) with
model KLFs of various ages (colored lines) for the Cloud 2-N cluster
(left) and Cloud 2-S cluster (right).
Red, green, blue, and orange lines represent the model KLFs of 0.1 Myr,
0.5 Myr, 1 Myr, and 2 Myr, respectively. The underlying IMF for the
model KLFs is the Trapezium IMF by \citet{Muench2002}.}
\label{fig:KLFmodel}
\end{figure}

\begin{figure}
\epsscale{0.9}
\plotone{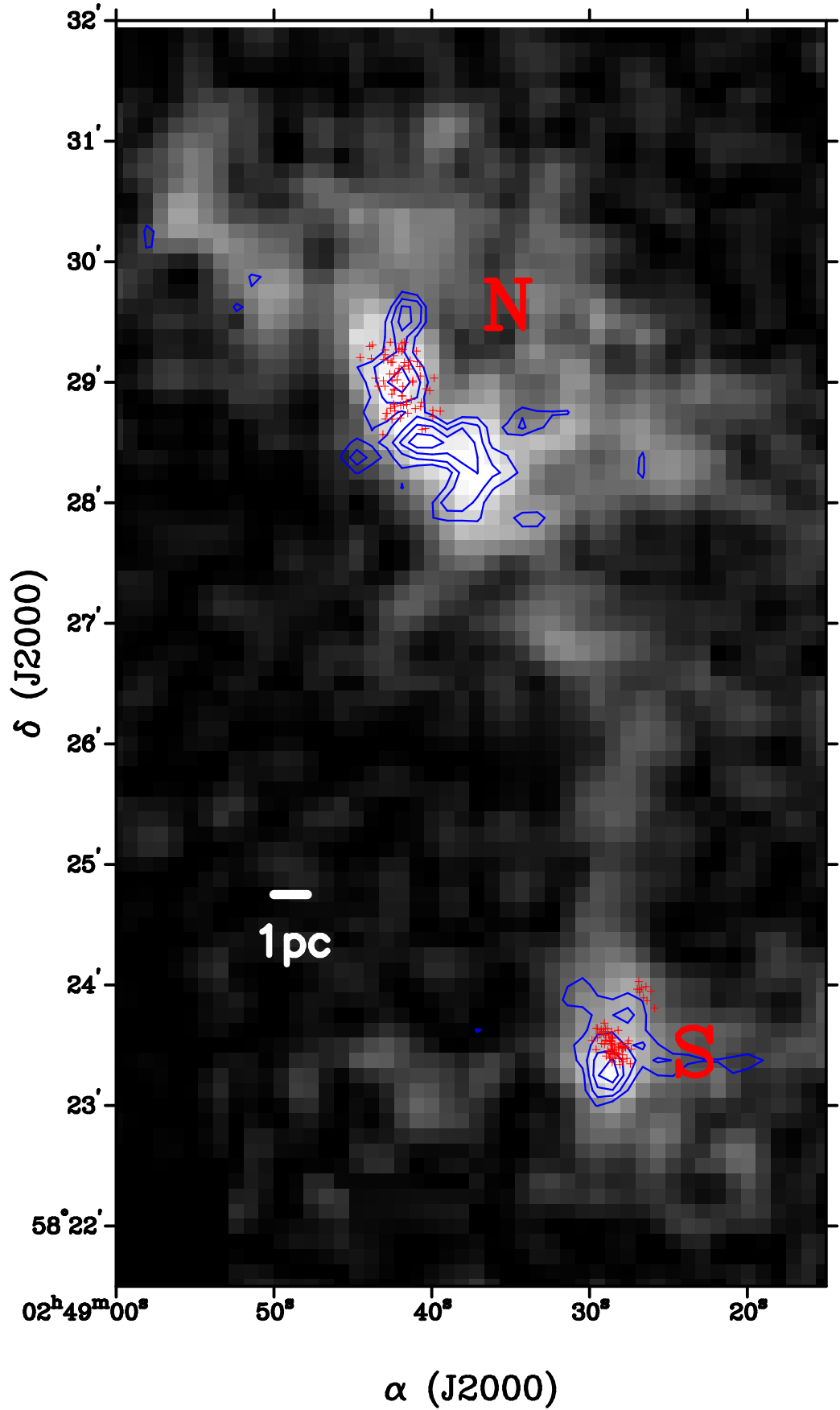}
%
%
\caption{Spatial distribution of Cloud 2 clusters (red crosses) and
Cloud 2 cores (blue contour for $^{13}$CO and gray scale for $^{12}$CO
from Saito et al. (2007)). The Cloud 2-N has two sub-peaks, while the
Cloud 2-S core appears to be single core.  The Cloud 2-N cluster seems
to be associated only with northern core of the Cloud 2-N. The contour
interval is 
%
0.05 K km s$^{-1}$ and ranges from 0.20 to 0.45 K km s$^{-1}$ 
and  the gray-scale range is from 0.0 to 1.7 K km s$^{-1}$. 
The temperature scale is in T$_{\rm A}^*$.
The beam size is about $19''$ for both $^{12}$CO and $^{13}$CO.}
\label{fig:CO_NIR} 
\end{figure}

\end{document}